\documentclass{article}
\usepackage[utf8]{inputenc}
\usepackage{authblk}
\usepackage{url}
\usepackage{graphicx}
\usepackage{float}
\usepackage{amsmath}

\title{Daylight Photometry of Bright Stars - Observations of Betelgeuse at Solar Conjunction}

\author[1]{Otmar Nickel}
\author[2]{Tom Calderwood}
\affil[1]{Zum Schollberg 11, 55129 Mainz, Germany; otmar.nickel@web.de}
\affil[2]{1184 NW Mt. Washington Drive, Bend, OR 97703; tjc@cantordust.net}
\date{November 2021}

\usepackage[authoryear]{natbib}
\setcitestyle{authoryear,open={},close={}}
\usepackage{graphicx}

\begin{document}

\maketitle
\noindent Published in JAAVSO Vol. 49, 2021
\\

\noindent \textbf{Subject Keywords}

AAVSO keywords = Photometry; Instrumentation; Variable Star Observing

ADS keywords = Photometry; CCD observation\\
\\
\noindent \textbf{Abstract}

\noindent Betelgeuse is an important variable star with many observations in the AAVSO database, but there is an annual gap of about four months where Betelgeuse is close to the sun and not observable at night. This gap could be filled with daylight observations. The star is bright enough to be imaged with small telescopes during the day, so photometry is possible when the sun is up. We present V band photometry of \(\alpha\) Ori taken with an amateur telescope equipped with an interline-transfer CCD camera and neutral density filter. These data compare favorably with contemporaneous nighttime photometry. The method used is a variation on ensemble photometry (using other bright daytime stars), and involves large stacks of very short exposures. The ensemble method provided V magnitudes of Betelgeuse with calculated errors of \(0.020\pm0.008\) mag from February to April 2021. From May to July, at the closest distances to the sun, the photometry of Betelgeuse could be continued with mean errors of \(0.040\pm0.013\) mag.

\newpage
\section{Introduction}
Betelgeuse has been a subject of great astrophysical interest. Ground-based telescopes can acquire reliable photometry from roughly early September to late April. That leaves four months without data in each year's light curve. Given the extraordinary recent behavior of this star, that break in coverage is most unfortunate. While space-based photometry of Betelgeuse is now practiced during solar conjunctions (\cite{STEREO}), the technique is difficult and cannot be performed in a standard photometric passband.

With care, it is possible to gather daytime aperture photometry of Betelgeuse. Such data were collected from February to July of 2021 with a small telescope located in Central Europe at only 200m elevation, where the atmospheric clarity is far from exceptional. 

\section{Observations and Data Acquisition}
The telescope used is of Newtonian configuration with aperture 250mm and focal length 1250mm, carried on a computer-controlled German equatorial mount. The optical tube is fitted with a cylindrical sun shade extending 50cm beyond the aperture. Together with a secondary shade around the focuser aperture, the shielding permits pointings as close as 10 degrees to the sun.

The camera is an ATIK model 460exm CCD, equipped with a wheel of photometric filters. For daylight work an additional neutral density (ND) filter of 1 percent transmission (as proposed by \cite{miles2007daytime}) is mounted ahead of the wheel. The camera is cooled to 0C for daytime work and has its own sun shade. Pointing of the mount is accurate enough to place a target star directly in the camera field of 20'x30'.

During photometry only a 687x550 pixel region of the CCD is read out. Image frames were typically integrated for 0.1 sec, though sometimes for 0.2 sec.  This kept the daytime sky to 50 percent or less of the sensor full-well depth. On each target, consecutive frames were taken to accumulate 10 sec of total exposure time. This process takes approximately 150 sec when using 0.1 sec exposures. Since guiding is not possible in daylight, the pointing is subject to drift. To mitigate this effect, it is helpful to use groups of ten exposures co-added and saved and later re-registered to make the 10 sec stack (this was done for about half of the presented photometry). Flat-field images were acquired by exposing against the daytime sky through a 3mm thick white polystyrene foam board. The benefit of the 10 sec stack is illustrated in figure 1. Depending upon the surrounding sky brightness, stars as dim as V~=6.5 can be successfully sampled this way. 

\begin{figure}[ht!]
\centering
\includegraphics[scale=0.45]{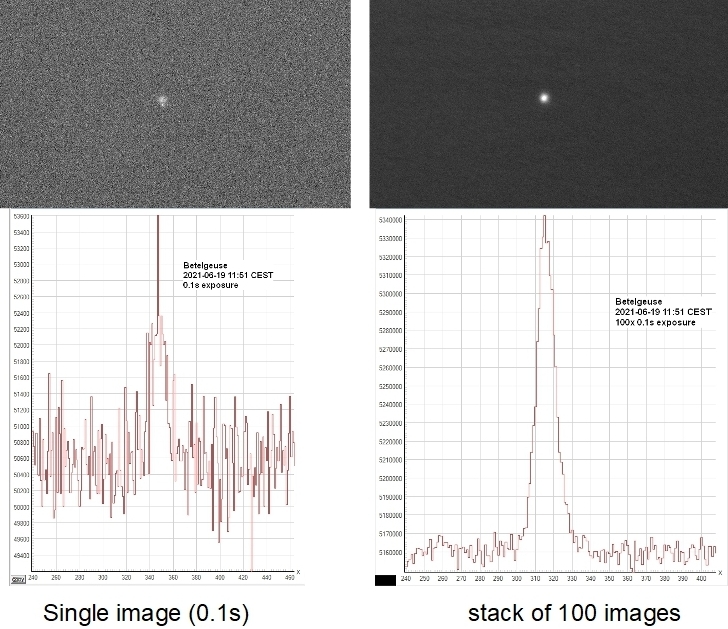}
\caption{images and profiles of Betelgeuse near to sun, left: single image, right: stack of 100 images}
\label{fig:Betelgeuse_profile}
\end{figure}

\newpage
\section{Sky background}
The brightness of the sky can be measured (in \(mag/arcsec^2\)) if a bright star with known magnitude is included in the field. The value is calculated by the formula:

\begin{equation}\label{skybrightness2}
m_{sky}=-2.5\log(\frac{N_{sky}} {p^2N_{star}})+m_{star}    
\end{equation}
where\\
\(N_{sky}\) = mean counts/pixel within sky annulus\\
\(N_{star}\) = summed pixel counts within star aperture after subtraction of sky background \\
p = pixel scale in arcsec/pixel

\begin{figure}[h]
\centering
\includegraphics[scale=0.28]{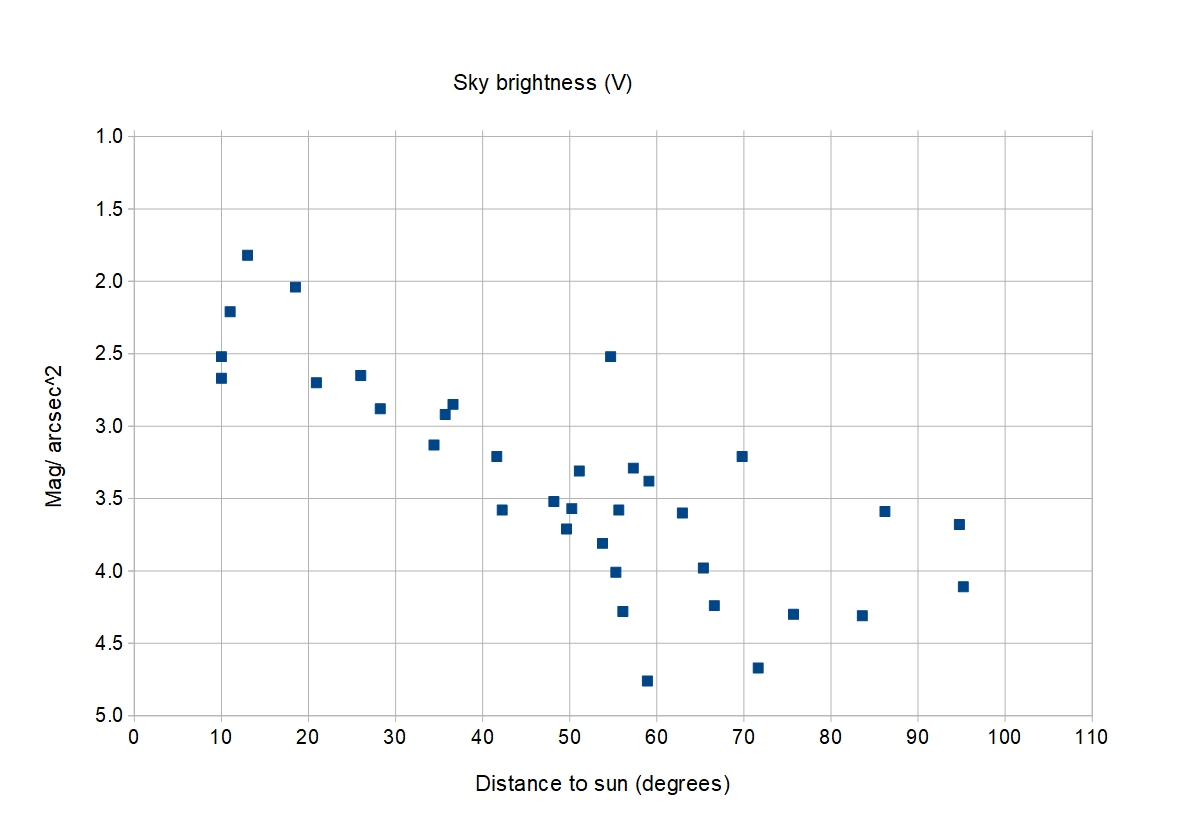}
\caption{Measured brightness of sky background in \(mag/arcsec^2\) versus distance from sun}
\label{fig:skybrightness}
\end{figure}

This method was first tested on the nighttime sky and gave values comparable to those of the “Sky Quality Meter” (SQM, \cite{Skyqualitymeter}), which is not usable in daylight. The individual measurements were repeatable within \(\pm0.1  mag/arcsec^2.\)

The daylight sky brightness depends on several factors, the prime one being angular distance of the field from the sun. As a test, the sky brightness was measured on 17 cloudless days (July to September 2020 and February to April 2021) with selections from 22 stars at different distances from the sun. The altitude of the sun ranged from 10° to 52°. The results are seen in fig.\ref{fig:skybrightness}. As expected, the smaller the angular distance from the sun, the higher the sky brightness.

\begin{figure}[ht!]
\centering
\includegraphics[scale=0.15]{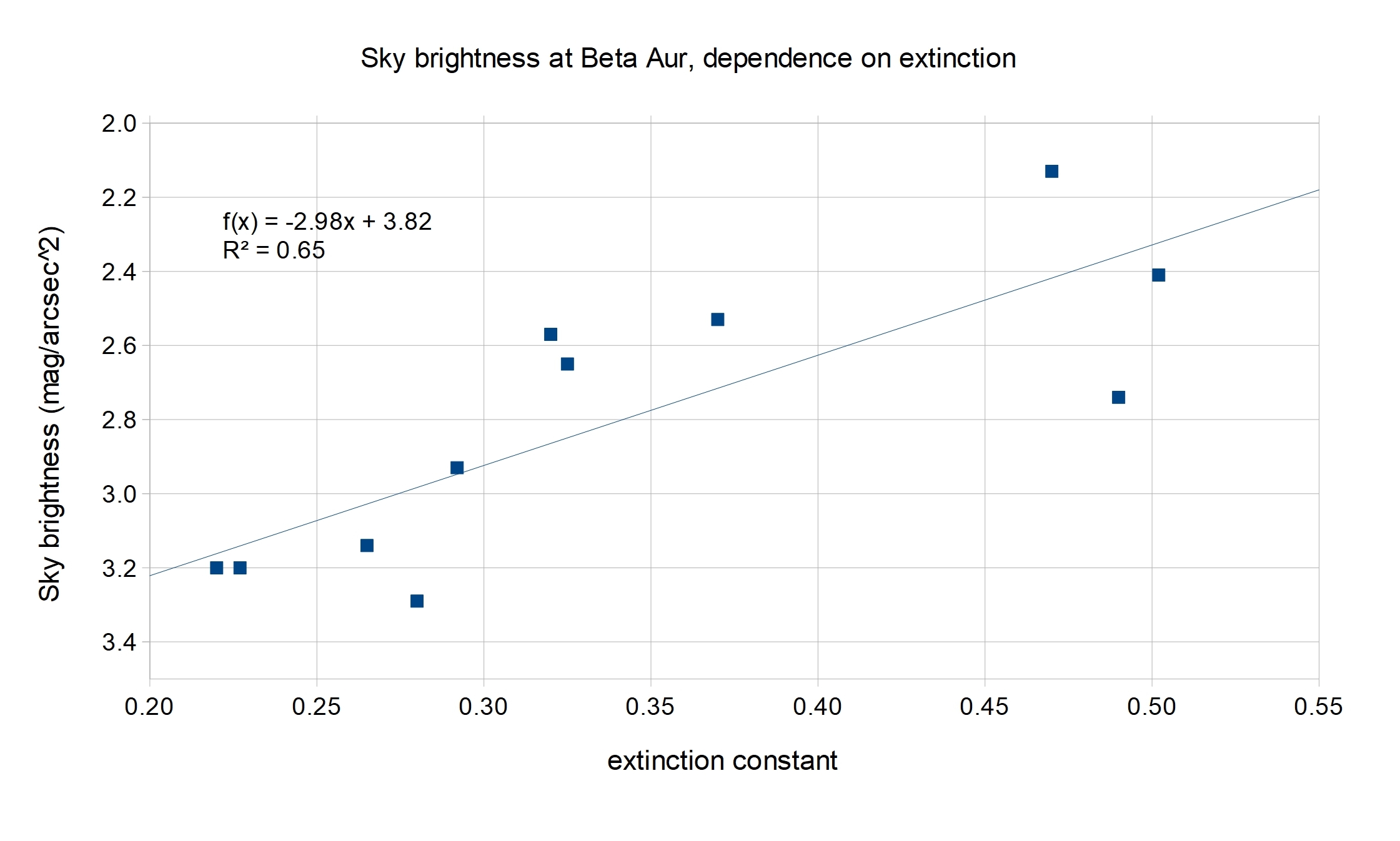}
\caption{Sky brightness around \(\beta\) Aur at noon versus extinction coefficient}
\label{fig:sky_extinction}
\end{figure}

Other factors are the altitude of the sun and the level of water vapor in the atmosphere. The latter is correlated to the extinction coefficient, so there is also a correlation between sky brightness and extinction, which can be seen in fig. \ref{fig:sky_extinction}. This figure shows the measured sky brightness around \(\beta\) Aur versus the measured extinction coefficient during 9 days between May 31 and July 22, 2021 around local solar noon.

Additional factors, e.g. airmass or the strong polarization of light at daytime can influence the sky brightness as well.

\section{Photometry methods}
\subsection{Instrumental and standardized magnitude}
Instrumental magnitudes \(m_{inst}\) were calculated using rectangular apertures for both star and sky background. The instrumental magnitude is given by:

\begin{equation}\label{instrumental magnitude}
   m_{inst} = -2.5 log(N_{star}/t) \\
\end{equation}
where

\(N_{star}\) = background-subtracted pixel counts within star aperture, t = exposure time in s

The processing of the images was done with the Fitsmag software package (\cite{FITSMAG}).

The standardized V magnitude of a star can be calculated from the instrumental magnitude by (\cite{da1992basic}):
\begin{equation}\label{daCosta}
   V = m_{inst} + a_0 + a_{1}(B-V) + a_{2}X + a_{3}X(B-V) + ...  
\end{equation}
where

(B-V) = color index of star

X = airmass of the observation

\(a_0\) = the zero point magnitude (in the following written as \(m_0\))

\(a_1\) = the color-term (in the following called Transformation coefficient \(T_v\) )

\(a_2\) = primary extinction coefficient (in the following replaced by \(-k_v\) )

\(a_3\) = second order extinction coefficient

\noindent The second order extinction coefficient is very small for the V band and can be neglected.

\begin{figure}[h]
\centering
\includegraphics[scale=0.14]{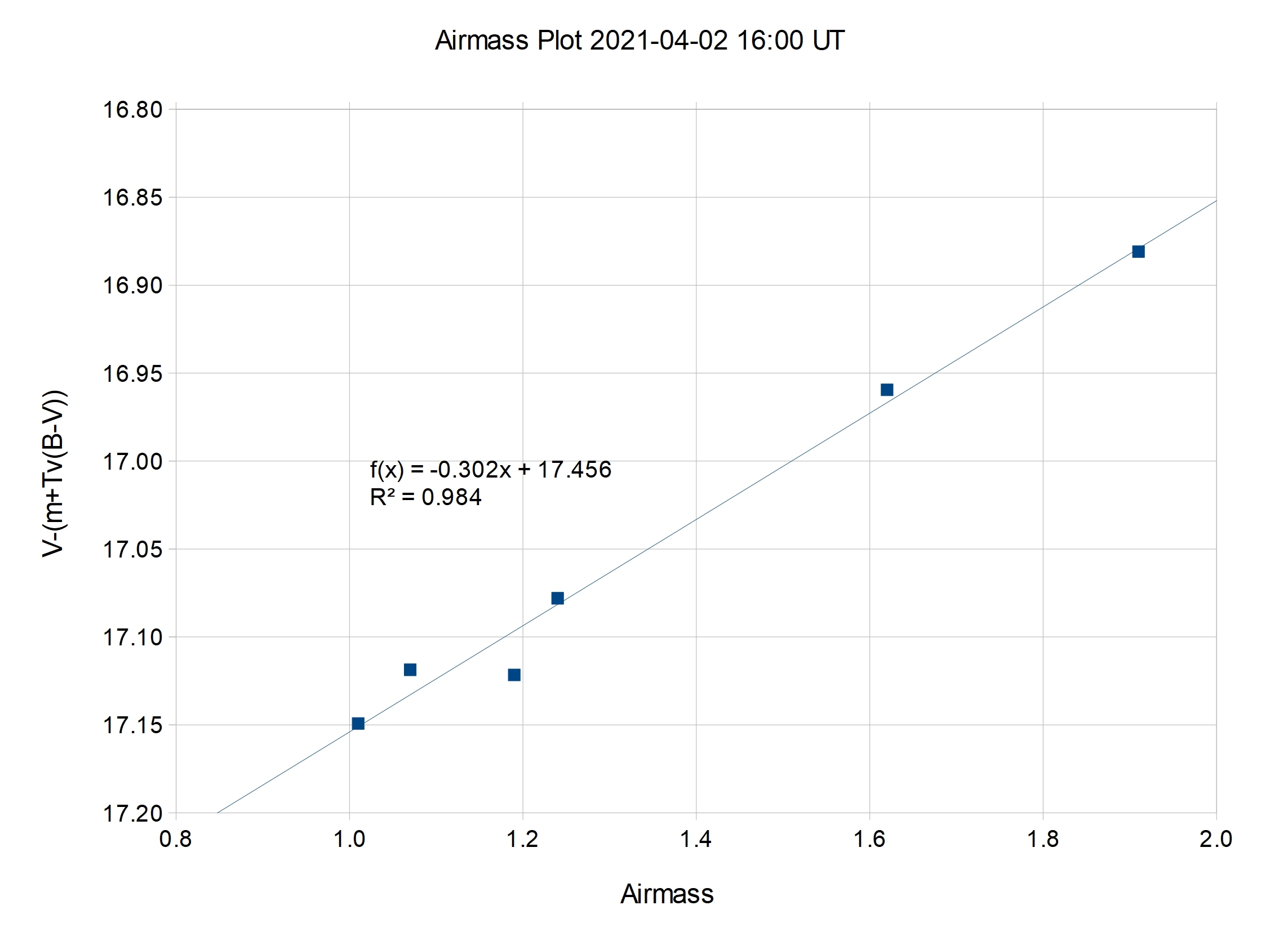}
\caption{Plot of \(V-(m_{inst}+T_{v}(B-V))\) vs. airmass of 6 bright stars with linear regression line}
\label{fig:airmassplot}
\end{figure}

Having in hand values for the transform coefficient and the target star's airmass, color, and instrumental magnitude, a standard magnitude can be calculated if the zero point and primary extinction coefficient can be established. This is done with observations of an ensemble of reference stars at a range of airmasses. For each of these stars, we have:
\begin{equation}
    V = m_{inst} +m_0 + T_v(B-V) - k_v X
\end{equation}
and
\begin{equation}\label{Airmass equation}
    V - (m_{inst} + T_v(B-V)) = m_0 - k_v X    
\end{equation}
where \(k_v\) and \(m_0\) are unknowns. If the differences \(V-(m_{inst}+T_{v}(B-V)\) of several stars are plotted against their airmasses , then \(k_v\) and \(m_0\) can be calculated by linear regression (LR) as shown in fig.\ref{fig:airmassplot}.\\

With known \(m_0\) and \(k_v\) the standard magnitude can be calculated by:
\begin{equation}\label{Ensemble equation}
    V_{var} = m_{inst} + T_{v}(B-V)_{var} + m_0 - k_{v}X_{var} 
\end{equation}
It can be shown (appendix A) that the result of this calculation is mathematically equivalent to differential ensemble photometry, where a mean value of differential magnitudes against an ensemble of comparison stars is calculated.

\begin{figure}[ht!]
\centering
\includegraphics[scale=0.14]{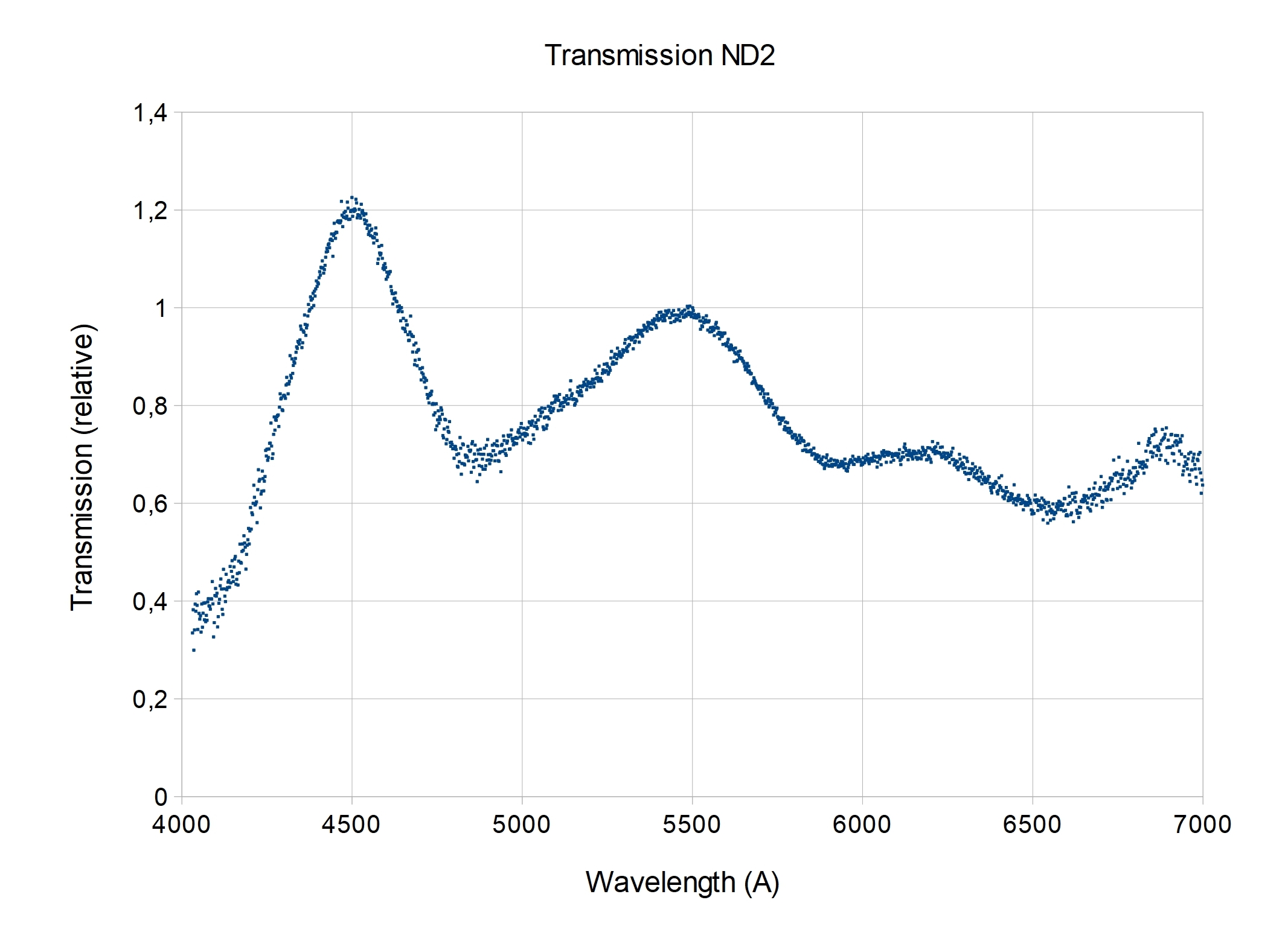}
\caption{Transmission curve of ND filter (units relative to 550nm)}
\label{fig:NDfiltercurve}
\end{figure}

 \(T_{v}\) for the combined V and ND filters was \(-0.0026 \pm 0.0057\). For the V filter alone it was +0.024, so the ND filter shifts the color significantly. The spectral transmission of the ND filter (Vendor: Antares) was analyzed by the author using a slit spectrograph measuring the fraction of transmitted light of a LED light source: it has a local peak at 550 nm, as shown from the transmission curve in fig. \ref{fig:NDfiltercurve}.

\subsection{Signal to noise ratio}
The well known sources of noise in CCD photometry are shot noise, thermal electronic noise, and readout noise. The quantum signal-to-noise-ratio (SNR) of an observation can be calculated using the "CCD equation" (\cite{merline1995}). In a high-count regime, this equation is approximated by:
\begin{equation}\label{ccd_equation}
SNR_q= \frac{N_{star}}{\sqrt{N_{star}+n_{pix}(1+\frac{n_{pix}}{n_{sky}})N_{sky}}} 
\end{equation}
where

\(N_{star}\) = number of collected photo electrons from star

\(n_{pix}, n_{sky}\) = number of pixels in star and sky apertures

\(N_{sky}\) = number of collected electrons/pixel from sky background\\
\\
The inverse of \(SNR_q\) may be called "normalized quantum noise" \(\sigma_{q}\) :
\[\sigma_{q}=\frac{1}{SNR_q}\]

In short exposures scintillation noise can be significant. The normalized scintillation noise $\sigma_s$ is the standard deviation of a series of star intensities divided by the mean value of the intensities (if no other noise would be involved).
The total noise \(\sigma_{total}\) and the SNR from both CCD noise \(\sigma_{q}\) and scintillation \(\sigma_s\) are:
\begin{equation}\label{total_noise}
\sigma_{total}=\sqrt{\sigma_{q}^2+\sigma_{s}^2},  \hspace{1cm} 
SNR=\sqrt{\frac{1}{\sigma_{q}^2+\sigma_s^2}}    
\end{equation}

Scintillation noise can be estimated with the modified "Young formula" (\citet{young1967photometric} and \citet{osborn2015atmospheric}):
\begin{equation}\label{scintillation_formula}
\sigma_s=0.003953\times X^{3/2}D^{-2/3}e^{-H/8000}\sqrt{t^{-1}}     
\end{equation}
where\\
D=aperture of telescope in m, H=local height above sea level in m, X=airmass, t=exposure time in s.\\
For the observing station (D=0.25m, H=200m) this gives the following values of \(\sigma_s\):\\
\begin{tabular}{c|c|c|c}
   & X=1.0 & X=1.5 & X=2.0\\
  t=1s & 0.00976 & 0.01793 & 0.0276\\
  t=10s & 0.00309 & 0.00567 & 0.00874\\
\end{tabular}
\\

Actual scintillation measurements (at daylight) were tested against the formula and demonstrate its reliability in this application.

\subsection{Total Photometric error}
The statistical error of ordinary differential photometry can be estimated as follows:
\begin{equation}
   error [mag] = \sqrt{error^2 (variable) + error^2 (comparison)} 
\end{equation}

In the ensemble approach used here, the zero point is the reference magnitude, and its error is a function of the LR:

\begin{equation}\label{ensemble_error}
  error[mag] = \sqrt{error^2(var) + (X_{var} -\overline{X})^2\sigma^2(k) +SER^2}  
\end{equation}
where\\ 
\(error(star) = 1.0857 \frac{1}{SNR(star)} \)\\
\(X_{var}\) = airmass of variable\\
\(\overline{X}\) = mean value of the airmass of the reference stars\\
\(\sigma(k)\) = error of extinction coefficient\\
SER = standard error of the LR (equation \ref{Airmass equation}).\\
This formula is derived in Appendix A.

\subsection{Reference stars for Betelgeuse}
The reference stars must be bright enough in comparison to the sky. The SNR should be not much below 50; this is the case if the star magnitude is nearly equal to the sky brightness (in \(mag/arcsec^2\)). From fig. \ref{fig:skybrightness}, it follows that down to a distance of 30° to the sun the star magnitude should be at least V=2.8; between 15° and 30° it should be brighter than 2.\\
Most bright stars have a slight variability, therefore only those with a magnitude range of less than 0.1 mag (from GCVS (\cite{GCVS})) were selected as comparison stars. V-magnitudes were taken from the Extended Hipparcos catalogue (\cite{anderson2012xhip}). Table \ref{tab:Betelgeusecompstars} shows the selected stars.\\
Depending on their distance from the sun, not all of these stars can always be used. Betelgeuse is closest to the sun on June 20, 2021, therefore the distance of the stars to the sun are listed for this date in the table. Around this date \(\beta\) Tau and \(\gamma\) Gem are too close to the sun; from June 30, \(\alpha\) and \(\gamma\) Gem cannot be used.\\
\(\alpha\) Gem has two components (A,B), separated by 5.5", which have to be measured together. If it is used in the ensemble, the aperture must be large; that could increase the error. \(\zeta\) Ori is a double star with 2.2" separation, which may be no problem.  \(\beta\) Aur is an EA variable with known period; it should be used only outside of the eclipses.

\begin{center}
\begin{table}
    \begin{tabular}{|c|c|c|c|c|c|c|}
        star & Vmag & B-V & \(\delta\)mag & dist. to  & sun dist.& comment  \\
        name & (XHip) & mag & (GCVS) & \(\alpha\) Ori & 2021-06-20 & \\
        & & & & & & \\
      \(\beta\) Ori & 0.18 & -0.03 & 0.05 & 19° & 33° & \\
      \(\alpha\) CMi & 0.40 & 0.43 & 0.07 & 26° & 32° & \\
      \(\beta\) Gem & 1.16 & 0.99 & 0.07 & 33° & 25° & \\
      \(\alpha\) Gem & 1.58 & 0.03 & 0 & 34° & 24° & double(5.5")\\
      \(\gamma\) Ori & 1.64 & -0.22 & 0.05 & 7.5° & 18° \\
     \(\beta\) Tau & 1.65 & -0.13 & 0 & 22° & 7° &\\ 
     \(\zeta\) Ori & 1.74 & -0.20 & 0.07 & 10° & 25° & double(2.2")\\
     \(\beta\) Aur & 1.90 & 0.08 & 0.09 & 37° & 21.5° & EA(P=3.96d)\\
     \(\gamma\) Gem & 1.93 & 0.00 & 0 & 14° & 12° &\\
     \(\alpha\) Ari & 2.01 & 1.15 & 0.06 & 57° & 51° &\\
     \(\alpha\) Cas & 2.24 & 1.17 & 0.07 & 78° & 64° &\\
    \end{tabular}
    \caption{\label{tab:Betelgeusecompstars}Reference stars for Betelgeuse}
    \end{table}
\end{center}

\section{Observation results}
\subsection{Results of reference stars and extinction}
From February 21 to July 31, 2021, daytime observations of Betelgeuse, \(\gamma\) Ori (as check star) and 4-7 reference stars were performed during 61 runs on 33 days. The reference stars were selected from among the following:
\(\alpha\) Gem, \(\alpha\) Ari, \(\alpha\) Cas, \(\beta\) Ori, \(\alpha\) CMi, \(\beta\) Gem, \(\beta\) Tau, \(\zeta\) Ori, \(\beta\) Aur, \(\gamma\) Gem.

\begin{figure}[ht!]
    \centering
    \includegraphics[scale=0.15]{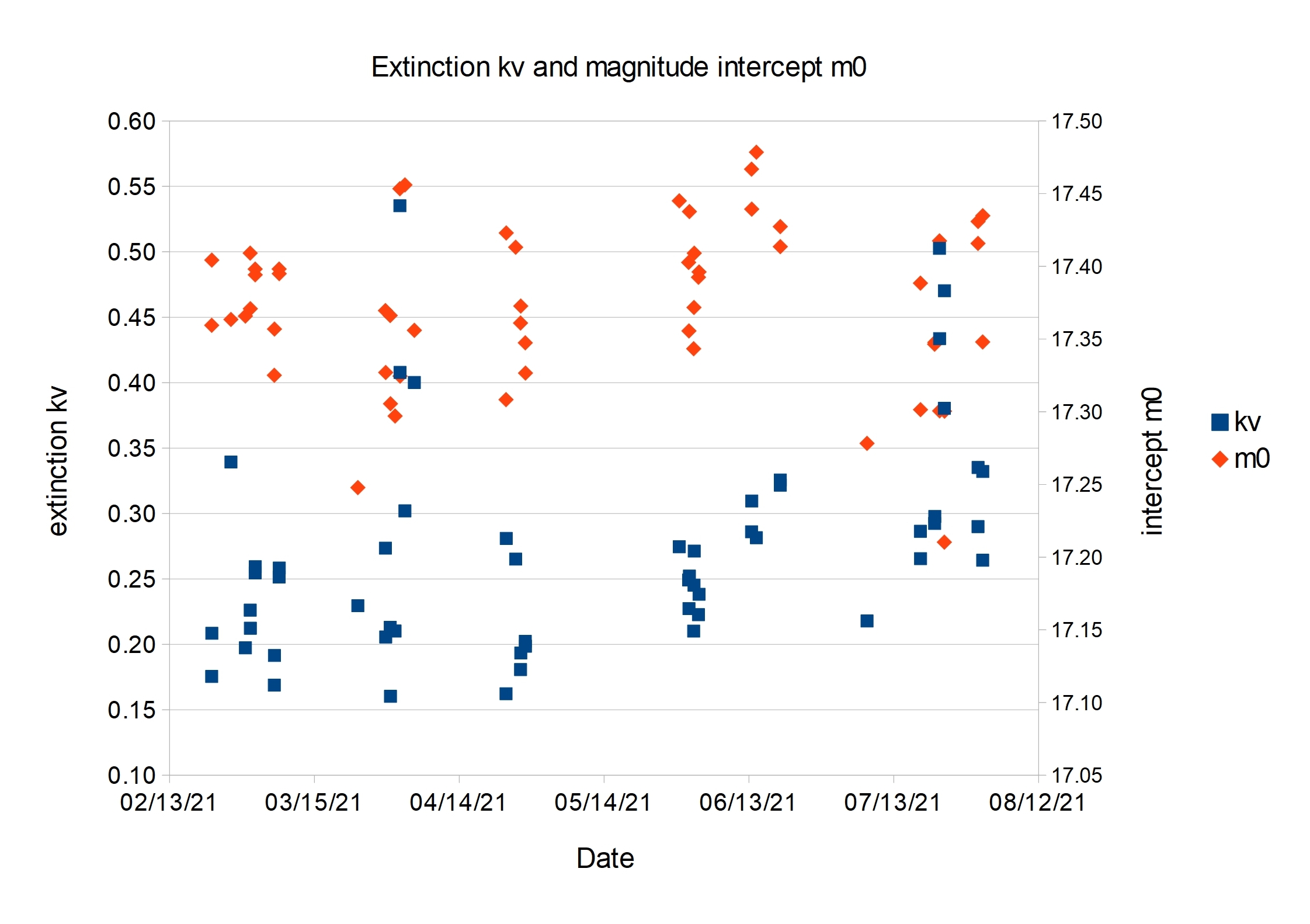}
    \caption{Values for extinction coefficient \(k_v\) (squares) and intercept \(m_0\) (diamonds)}
    \label{fig:extinction_m0}
\end{figure}

The SNR of the star measurements used was typically above or around 100. Results with SNR below 50 (2 of 200) were discarded. The remaining range was between 53 and 290. The LR statistics for Betelgeuse observations are shown in table \ref{tab:linregresults}. The first group represents data from February to April 2021, the second, May to July 2021, where the stars are closer to the sun. The errors of the second group are significantly higher, as can be expected from the brighter sky and from the worse seeing during this time. The extinction was also higher due to many hazy days. All values of \(k_v\) and \(m_0\) are shown in fig. \ref{fig:extinction_m0}.\\
\\
The whole range of extinction values was between 0.16 and 0.53; values above 0.3 corresponded to a very hazy sky. The standard error did not correlate significantly to the extinction value.\\

For the LR, up to one star was excluded if the deviation of the star magnitude from the regression line was significant (\(>2\times std.error \)) and greater than 0.05 mag. In 4 of 65 regression calculations this rule was used.

\begin{table}
\begin{tabular}{c|c|c}
     & February to April & May to July\\
     & mean value (std. dev.)& mean value  (std. dev.) \\
standard error & 0.017 (0.008) & 0.036 (0.011)\\
extinction constant \(k_v\) & 0.247 (0.084) & 0.299 (0.073)\\
error of \(k_v\) & 0.027 (0.010) & 0.045 (0.016)\\
intercept \(m_0\) & 17.365 (0.047) & 17.381 (0.064)\\
\(R^2\) & 0.962 (0.036) & 0.917 (0.056)\\
\end{tabular}
    \caption{\label{tab:linregresults}Statistics of linear regression from February to July 2021}
\end{table}

\subsection{Observations of Betelgeuse (February to July 2021)}

\begin{figure}[htbp]
    \centering
    \includegraphics[scale=0.14]{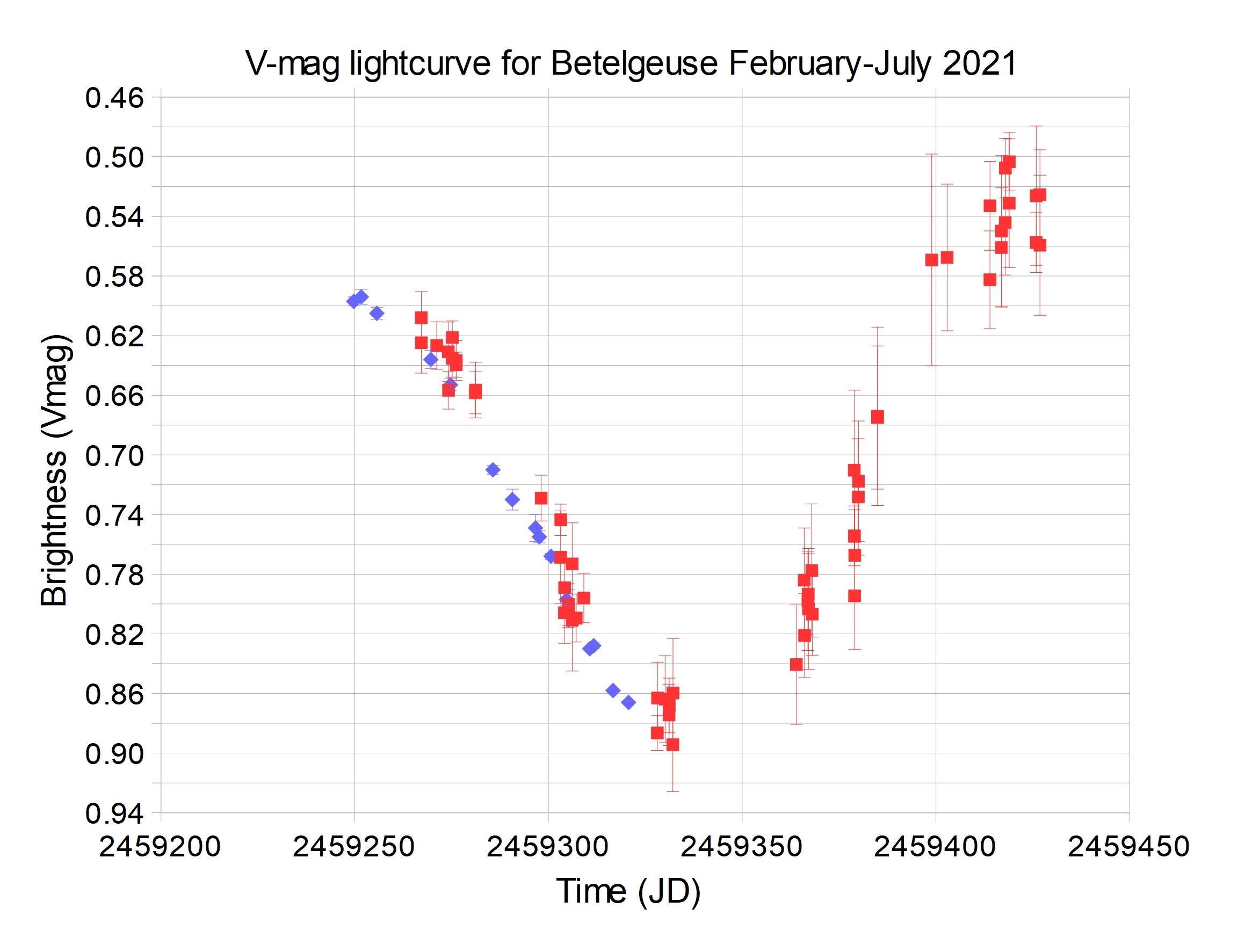}
    \caption{Lightcurve (V-magnitudes) of Betelgeuse February to July, 2021. Blue diamonds: PEP data (night), Red squares: CCD data (daylight). PEP error bars are too small to see.}
    \label{fig:Betelgeuse2021}
\end{figure}

The daylight observations of Betelgeuse from February, 21 to July, 31 (2021) together with nighttime photoelectric photometry (PEP) are shown in fig. \ref{fig:Betelgeuse2021}. The PEP data were collected with an Optec SSP-3 photometer mounted on a 235mm telescope in North America. The PEP V data were taken in concert with B data and the \(\Delta\)(B-V) with respect to the comparison star (HD 37160) was established from the instrumental magnitudes and transformation coefficients. During the period 2459255 to 2459327 the mean \(\Delta\)(B-V) was 0.910 with a standard deviation of 0.010. The comparison (B-V) is taken as 0.958, implying that the mean (B-V) of Betelgeuse was 1.868. The color transformation of the daylight measurements were based on a constant (B-V) of Betelgeuse of 1.85 (from Hipparcos).
 
\begin{figure}[ht!]
    \centering
    \includegraphics[scale=0.12]{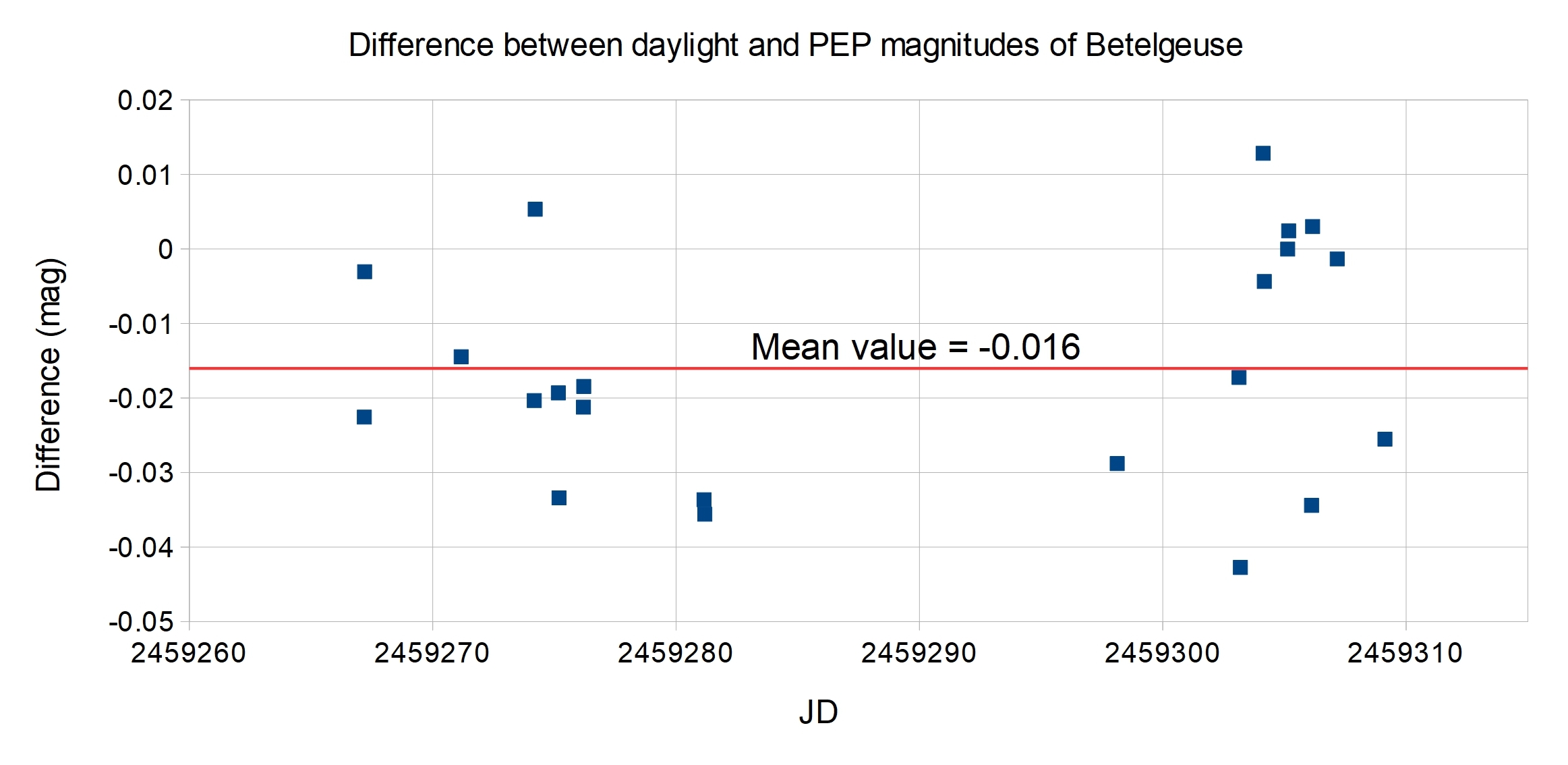}
    \caption{Difference between daylight and interpolated PEP magnitudes of Betelgeuse. Horizontal line: mean value.}
    \label{fig:PEPdiff}
\end{figure}

The PEP magnitudes are in most cases slightly dimmer and the difference between the daylight results and interpolated PEP magnitudes is shown in fig. \ref{fig:PEPdiff}. The mean value of the differences is \(-0.016\pm0.003\) mag (n=22). The solar angle was between 16° and 114° and the sky brightness around Betelgeuse was between 1.8 and 6.0 \(mag/arcsec^2\).

The mean errors of the daylight results were 0.02 mag (February-April) and 0.04 mag (May-July), which is only slightly higher than the LR standard errors of the linear regression of the reference stars (table \ref{tab:linregresults}); this is a consequence of the instrumental magnitude error of Betelgeuse in the range 0.006-0.008 mag, which was much smaller than the standard error in all cases, as well as the extinction error (airmass term in equ. \ref{ensemble_error}), which was in the range 0.001-0.015 mag.

\begin{figure}[ht!]
    \centering
    \includegraphics[scale=0.12]{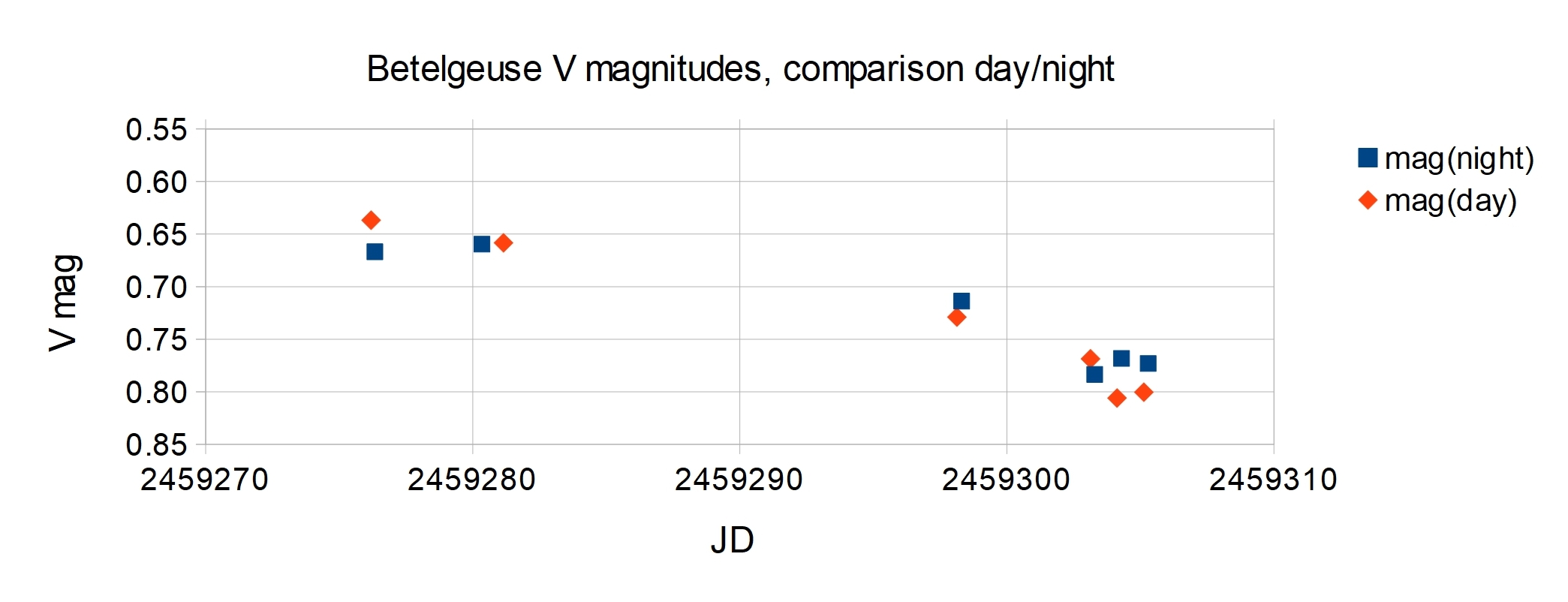}
    \caption{Comparison of daytime and nighttime magnitudes of Betelgeuse resulting from the same method. Blue squares: Nighttime results; Red diamonds: Daytime results}
    \label{fig:day_night}
\end{figure}

In 6 nights, from March 02 to March 31, 2021, Betelgeuse was observed also at night with the same CCD equipment and method as at daytime, and the magnitudes were calculated using the same comparison star ensemble as at daytime. The results are shown in fig. \ref{fig:day_night}. The mean values of the differences between day and night magnitudes are shown in the following table:

\begin{tabular}{c|c|c|c}
     & Difference (day-night) & error(day) & error(night)  \\
  mean value (n=6)   & 0.006 mag & 0.018 mag & 0.020 mag\\
  std. dev.    & 0.026 & 0.007 & 0.012\\
  error of mean & 0.011 & 0.003 & 0.005 \\
\end{tabular}

Day and night magnitudes are not significantly different; the mean values of the calculated errors are also in the same range.\\

All daylight observations were uploaded to the AAVSO International Database, where the PEP data may also be found (www.aavso.org/LCGv2). If more than one daylight observation was available per day, only the mean value of the magnitudes was recorded.

\section{Discussion}
The accuracy of the magnitude results depends partly on the selection of reference stars. For consistency, the magnitudes of the Betelgeuse ensemble stars were all taken from the Hipparcos catalogue. There are other catalogues, e.g. the General Catalog of Photometric Data (GCPD, \cite{GCPD1996}). For some stars used as references, there are differences between Hipparcos and GCPD of up to 0.042 mag. In a sample of 3 cases, where \(\beta\) Ori, \(\alpha\) CMi and \(\beta\) Gem were used, the Vmag result with GCPD values differed by about -0.01 mag and the error was slightly smaller. Therefore future projects should use the GCPD magnitudes of the reference stars, if available.

The systematic differences of -0.016 (\(\pm 0.003\)) between daylight and PEP magnitudes cannot result from differences in the (B-V) values used for Betelgeuse in transformation: the PEP and daylight (B-V) differ by only about 0.018. This would result in an offset of less than 1 mmag in the daylight photometry. But the difference can be explained by the error of the color transformation constant \(T_v\). Because of the high color index of Betelgeuse of around 1.85 this difference can be induced by a \(T_v\) error of -0.0097 (mean value of color indices of the reference stars was around 0.2 mag). The calculated \(T_v\) error was only slightly lower (0.0057), therefore this explanation is plausible. If also the GCPD magnitudes would be used, the systematic error would be in the order of -0.026 and the error of \(T_v\) would be in the order of 0.016. Therefore in future projects the transformation constant should be adjusted in some way.

\section{Conclusions}
We have shown that CCD photometry in daylight with amateur equipment can be performed with an accuracy in the order of 0.02-0.05 mag. This is possible even within 10° of the sun, at least for stars of brightness on the order of V\(<=\)2. Daylight photometry of Betelgeuse has been compared to PEP photometry at night during 8 weeks; systematic differences between the lightcurves were less than 0.02 mag. Observations of bright variable stars could be collected in this way nearly uninterrupted by conjunctions with the sun.\\
\\
The photometry could be improved in three ways:\\
\\
(1) A camera with higher framing rates could enable stacking of more than 1000 0.1 sec frames.  The additional frames would increase the SNR, reduce the effect of scintillation, and allow the observation of stars with magnitudes in the range up to V=3.\\
\\
(2) Observations in regions with altitudes above 1000 m and with low humidity may provide a darker sky and would also enlarge the range of observable stars.\\
\\
(3) Because the sky brightness is lower at longer wavelengths, it could be better to use redder filters to reduce the background noise, but in this case the transformation to the V magnitude would be more difficult. In any case an additional measurement with photometric R filter would give more information.

\bibliography{references}

\appendix
\section{Appendix: Ensemble comparison and linear regression}
%Appendix A: Ensemble comparison and linear regression

With LR, the values of  \(V_i - (m_i + T_v Y_i)\) of a sample of n reference stars with catalogue magnitudes \(V_i\), color index \(Y_i\) and instrumental magnitudes \(m_i\) are correlated to their airmass \(X_i\) via the linear function:
\begin{equation}\label{Refstar regression}
    V_i - (m_i + T_v Y_i) = m_0 - k_v X_i    
\end{equation}
LR of this formula provides the regression parameters \(m_0\) and \(k_v\); then for each reference star follows with the residual \(\epsilon_i\) the equation:
\begin{equation}\label{Refstar equation}
    V_i = (m_i + T_v Y_i) + \epsilon_i + m_0 - k_v X_i    
\end{equation}

The differential magnitude \(V_{var}\) of a variable with instrumental magnitude \(m_{var}\), with respect to a reference star (with instrumental magnitude \(m_i\)) would be:
\begin{equation}
    V_{var} = m_{var} - m_i + T_v(Y_{var}-Y_i) -k_v (X_{var}-X_i) + V_i
\end{equation}
The ensemble method uses the arithmetic mean value of all reference stars, therefore:
\begin{equation}
    V_{var} = \frac{1}{n} \sum_{i=1}^{n} (m_{var} - m_i + T_v(Y_{var}-Y_i) -k_v (X_{var}-X_i) + V_i)
\end{equation}
which simplifies to:
\begin{equation}\label{ensemble diff}
    V_{var} = m_{var} + T_v Y_{var} - k_v X_{var} - \frac{1}{n} \sum_{i=1}^{n} (m_i + T_v Y_i - k_v X_i - V_i)
\end{equation}
Now \(V_i\) is replaced by the right side of equation \ref{Refstar equation}:
\begin{equation}
\begin{split}
   V_{var} & =  m_{var} + T_v Y_{var} - k_v X_{var}\\
   & - \frac{1}{n} \sum_{i=1}^{n} (m_i + T_v Y_i - k_v X_i - (m_i + T_v Y_i) - \epsilon_i - m_0 + k_v X_i)
\end{split}
\end{equation}
which becomes:
\begin{equation}
   V_{var} = m_{var} + T_v Y_{var} - k_v X_{var} + \frac{1}{n} \sum_{i=1}^{n} (\epsilon_i + m_0)
\end{equation}
Because the sum of the residuals \(\epsilon_i\) is zero, this yields:
\begin{equation}
   V_{var} = m_{var} + T_v Y_{var} - k_v X_{var} + m_0
\end{equation}
This is exactly the same as equation \ref{Ensemble equation}. This means that the LR method is equivalent to a differential comparison method with an ensemble of comparison stars. Therefore the error calculation can be based on the mean value of differential magnitudes.

For the evaluation of the resulting error one works with equation \ref{ensemble diff}. This can be transformed to:
\begin{equation}
V_{var} = m_{var} + T_v (Y_{var}-\frac{1}{n}\sum_{i=1}^{n}Y_i)  -  k_v (X_{var} -\frac{1}{n}\sum_{i=1}^{n}X_i) - \frac{1}{n} \sum_{i=1}^{n} (m_i - V_i)
\end{equation}
where (\(\overline{X}\), \(\overline{Y}\) and  \(\overline{V}\) are mean values of \(X_i\), \(Y_i\) and \(V_i\))
\begin{equation}
V_{var} = m_{var} + T_v (Y_{var}-\overline{Y})  -  k_v (X_{var} -\overline{X}) - \frac{1}{n} \sum_{i=1}^{n}m_i + \overline{V}
\end{equation}
In this formula \(\overline{V}\) is constant,  \(m_{var}\), \(T_v\), \(Y_{var}\), \(k_v\) and \(m_i\) have statistical errors, which are independent of one another. Therefore \(\sigma^2(V_{var})\), the squared error of \(V_{var}\) can now be estimated from a sum of the variances of these errors:
\begin{equation}\label{ensemble_variance}
\begin{split}
\sigma^2(V_{var}) = & \sigma^2(m_{var}) + T_v^2\sigma^2(Y_{var})+ (Y_{var}-\overline{Y})^2\sigma^2(T_v)\\
& + (X_{var} -\overline{X})^2\sigma^2(k_v) +\frac{1}{n}\sum_{i=1}^{n} \sigma^2(m_i)
\end{split}
\end{equation}
The variance of \(m_i + T_v Y_i\) is obtained from the sum of the residuals of the linear regression divided by 1/(n-2):
\begin{equation}
\sigma^2(m_i + T_v Y_i) = \frac{1}{n-2}\sum_{i=1}^{n}\epsilon_i^2 
\end{equation}
The right side of the above equation is the square of the standard error of the regression (SER). It follows:
\begin{equation}
\sigma^2(m_i) =  SER^2 - \sigma^2(T_v Y_i) 
\end{equation}
and 
\begin{equation}
 \frac{1}{n}\sum_{i=1}^{n} \sigma^2m_i = SER^2 - \frac{1}{n}\sum_{i=1}^{n}Y_i^2\sigma^2(T_v) = SER^2 - \sigma^2(T_v)\frac{1}{n}\sum_{i=1}^{n}Y_i^2   
\end{equation}
From equation \ref{ensemble_variance} follows:
\begin{equation}\label{error_ensemble}
\begin{split}
\sigma^2(V_{var})  =  \sigma^2(m_{var}) + T_v^2\sigma^2(Y_{var})+ (Y_{var}-\overline{Y})^2\sigma^2(T_v)\\
 + (X_{var} -\overline{X})^2\sigma^2(k)
 + SER^2 - \sigma^2(T_v)\frac{1}{n}\sum_{i=1}^{n}Y_i^2  
\end{split}
\end{equation}
If the error of \(T_v\) is neglected, then the approximate error of V(ensemble) can be calculated by:
\begin{equation}\label{ensemble_err2}
  \sigma(V_{var}) = \sqrt{\sigma^2(m_{var}) + (X_{var} -\overline{X})^2\sigma^2(k) +SER^2}  
\end{equation}

\end{document}